\documentclass[twocolumn,amsmath,amssymb,pra,superscriptaddress]{revtex4}

\usepackage{graphics}
\usepackage{epsfig}
\usepackage{amsmath}
\usepackage{amsfonts}
\usepackage{color}
\usepackage{dcolumn}
\usepackage{bm}
\usepackage{mathrsfs}

\begin{document}

  \bibliographystyle{apsrev}
  
\title{Alternative Multipole Expansion of the Electron Correlation Term}

      \author{Eric Ouma Jobunga}
       \affiliation{\it  Department  of Mathematics and Physics , Technical University   of Mombasa,\\ P. O. Box 90420-80100, Mombasa, Kenya}
       \affiliation{\it  AG Moderne Optik, Institut f\"ur Physik,
         Humboldt-Universit\"at zu Berlin, Newtonstr.~15,
         12\,489 Berlin, Germany}
%
%


\begin{abstract}

An alternative multipole expansion of the correlation term is derived. Modified spherical Bessel type functions which simplify as a summation of multiple orders of basic trigonometric functions are generated from this new method.  We use this new expansion to obtain useful insights into the electron-electron interaction. An analytic expression for the electronic correlation term is suggested. Also, a pseudopotential for helium-like system is derived from this alternative expansion, and some reasonable eigenvalues for the ground state and two autoionizing levels of helium atom, is provided as a test of efficiency of this solution approach. With some additional corrections beyond the non-relativistic limit, a helium atom groundstate energy of $-2.9036$ is obtained using the analytical form derived from this method and the Slater determinant expansion of the wavefunction.

\end{abstract}

\maketitle
\section{Introduction}

Helium atom and helium-like ions are the simplest many-body systems containing two electrons which interact among themselves in addition to their interaction with the nucleus. The two-electron systems are therefore the ideal candidates for studying the  electron correlation effects.

The non-relativistic Hamiltonian of a two-electron system with a nuclear charge $Z$ is given by

    \begin{equation}
    \mathrm{H} = \frac{1}{2}\, \left[p_1^2 + p_2^2\right] - Z\, \left[\frac{1}{r_1} + \frac{1}{r_2} \right] + \frac{1}{|\mathbf{r}_1-\mathbf{r}_2|}
    \end{equation}
where the first term correspond to the sum of the kinetic energy of each of the two electrons, the second term to the sum of the interactions between each of the electrons and the nucleus, and the last term to the electron correlation interaction between the two electrons. The second and the last term form the potential energy function of a bound two-electron system.

 If the Hamiltonian is used to solve the time-independent Schr\"{o}dinger equation
    \begin{equation}
      \mathrm{H} \Phi_n(\mathbf{r}_1, \mathbf{r}_2) = E_n  \Phi_n(\mathbf{r}_1, \mathbf{r}_2)
    \end{equation}
 for any eigenstate $\Phi_n(\mathbf{r}_1, \mathbf{r}_2)$ of the system, the eigenenergies $E_n$ for the particular state are obtained. The major problem in many-body systems is the correlation term,  coupled with the fact that the wavefunction of the system is never exactly known, which complicates the reduction of the Schr\"odinger equation of the many-body system to a single-particle equation. This makes  the solution to the eigenvalue problem difficult. One has to therefore rely on some approximation methods in trying solve such a problem in order to obtain the correct eigenenergies and eigenvectors which may be useful for further estimation of many physical parameters like transition matrices, expectation values, polarizabilities and many others. 
 
 Difficult theoretical approaches have been used in the past in dealing with the electron correlation problem. Some of these approaches include the  variational Hyleraas method \cite{Hylleraas1929, Drake1999}, coupled channels method \cite{Barna2003}, the configuration interaction method \cite{Hasbani2000}, explicitly correlated basis and complex scaling method \cite{Scrinzi1998}. At present only the Hylleraas method, which includes the interelectronic distance as an additional free co-ordinate, yields the known absolute accuracy of the groundstate energy of the helium atom \cite{Pekeris1959}. Configuration interaction methods have also been proved to be accurate but they are quite expensive computationally. To overcome this computational challenge especially for really large systems, single active electron (SAE) methods become advantageous but they also require some approximations in developing the model potentials \cite{Parker2000, Parker1998} which can further be used to generate the eigenvectors and energies.  The development of the SAE models has become an active field of study taking different approximations \cite{Chang1976} like the independent particle approximation (IPA), multi-configurational Hartree-Fock (HF) \cite{Szabo1996}, density functional theory (DFT) \cite{Kohn1965}, random phase approximation (RPA) \cite{Linderberg1980}, and many others . The major limitation of SAE approximations is the inability to explain multiple electron features like double excitation, simultaneous excitation and ionization, double ionization but progress is being made towards the realization of these features. 
 
In this paper, an alternative multipole expansion is proposed. Based on this expansion, new modified spherical Bessel type functions are generated. In addition, we suggest an analytic expression
 \begin{equation}
 \begin{split}
 \frac{1}{|\mathbf{r}_1-\mathbf{r}_2|} &= {\frac{1}{\sqrt{r_1^2 + r_2^2}}\,\exp{\left\lbrace \frac{\mathbf{r}_1\cdot\mathbf{r}_2}{r_1^2 + r_2^2} \right\rbrace}} \label{eq:ct0}
 \end{split}
\end{equation}  
to describe the electron-electron interaction term.

 \section{The Alternative Multipole Expansion}
 
 The correlation term   can be written as
  \begin{equation}
    \frac{1}{|\mathbf{r}_1-\mathbf{r}_2|} = \frac{1}{r_>}\, (1 -2tx + t^2)^{-1/2} \label{eq:ct1}
    \end{equation}
 where ${t= \frac{r_<}{r_>}}$,  $r_<(_>)$ corresponds to the lesser (greater) electronic radial distance between the two electrons. In Legendre polynomials, equation (\ref{eq:ct1}) is  conventionally expressed as \cite{amo:BetheandSalpeter1957}
    \begin{equation}
    \frac{1}{|\mathbf{r}_1-\mathbf{r}_2|} = \sum_{l=0}^{\infty} \frac{r_<^l}{r_>^{l+1}} P_l(\cos \theta) \label{eq:ct2}
    \end{equation}
 where $P_l(\cos \theta)$ are the Legendre Polynomials of order $l$,  and $\theta$ is the relative azimuthal angle between the electron position vectors. In the alternative framework, the correlated term 
 \begin{equation}
  \begin{split}
    (1 -2tx + t^2)^n &= \sum_{s=0}^{\infty} \left(\begin{matrix} n \\ s \end{matrix} \right)\, (y_0(t))^{n-s} (y_1(t))^{s}\,x^s\\
    &= (y_0(t))^{n}\,\sum_{s=0}^{\infty} \left(\begin{matrix} n \\ s \end{matrix} \right)\,  \left(\frac{y_1(t)}{y_0(x)}\right)^{s}\,x^s  \label{eq:ct3}
    \end{split}    
    \end{equation}
  in equation (\ref{eq:ct1}) is expressed in a binomial expansion, similar to Gegenbauer polynomial \cite{cp:Gregory2011, Abramowitz1965} with ${n=-1/2}$, and the functions ${y_0(t)= 1 + t^2}$ and ${y_1(t)= -2\, t}$ defined. Ideally, this is the point of departure with equation (\ref{eq:ct2}) where the expansion of the correlated term is done as a summation of functions of $t^s$ with ${s\geq 0}$ as the summation index. The next step involves re-writing the expansion 
  \begin{equation}
  \begin{split}
    (1 -2tx + t^2)^n = (y_0(t))^{n}\,& \sum_{l=0}^{\infty}\,\sum_{s=0}^{\infty}\,  \beta_s^{(\hat{l}/2)}  \left(\begin{matrix} n \\ 2s + l \end{matrix} \right) \\ &\times  \left(\frac{y_1(t)}{y_0(x)}\right)^{2s+l}\,P_l(x) \label{eq:ct4}
    \end{split}    
    \end{equation}
  with $x^s$ as a function of the Legendre polynomials whose symmetry relations are of practical significance in the simplification of integrals using spherical co-ordinates. The coefficients $\beta_s^{(\hat{l}/2)}$ have an intrinsic connection between the index $s$ of $x^s$ and the Legendre polynomials $P_l(x)$, and ${\hat{l} = l}$ for even $l$ and ${\hat{l} = l-1}$ for odd $l$. The exact recursive pattern for the coefficients $\beta_s^{\hat{l}/2}$ is subject to further investigation. Below, we present the pattern 
  \begin{equation}
  \begin{split}
  \beta_s^{(0)} &= \frac{(2l+1)}{(2l+2s+1)}\\
  \beta_s^{(1)} &= \frac{(2l+1)\,2^1\,(s+1)}{(2l+2s+1)\,(2l+2s-1)}\\
  \beta_s^{(2)} &= \frac{(2l+1)\,2^2(s+1)\,(s+2)}{(2l+2s+1)(2l+2s-1)(2l+2s-3)}\\
   \vdots\\
   \beta_s^{(k)} &= \frac{(2l+1)\,2^k(s+k)!\,(2l+2s-(2k+3))!!}{(2l+2s+1)!!\,s!}  \label{eq:ct5}
  \end{split}
  \end{equation}
corresponding to ${l \leq 5}$ but generalized for all $l$ values. Substituting ${n=-1/2}$ and the variables $y_0(t)$ and $y_1(t)$ into equation (\ref{eq:ct4}) and simplifying leads to
  \begin{equation}
  \begin{split}
    \frac{1}{\sqrt{1 -2tx + t^2}} =  & \frac{1}{\sqrt{1+t^2}}\,\sum_{l=0}^{\infty}\,\sum_{s=0}^{\infty}\, \beta_s^{(\hat{l}/2)}  \frac{(2l + 4s+1)!!}{(2s+l)!}\\ & \times \left(\frac{t}{1+t^2}\right)^{2s+l}\,P_l(x). \label{eq:ct6}
    \end{split}    
    \end{equation}
 The correlation interaction in equation (\ref{eq:ct1}) can be expressed as a multipole summation series
  \begin{equation}
  \begin{split}
    \frac{1}{r_1\, \sqrt{1 -2tx + t^2}} = \frac{4\,\pi}{r_1\,\sqrt{1+t^2}}\sum_{l,m=0}^{\infty}\,\tilde{j}_l(t)\, Y_l^{m*}(\hat{r}_1)Y_l^{m}(\hat{r}_2). \label{eq:ct7}
    \end{split}    
    \end{equation}
 where $Y_l^{m}$ are the spherical harmonics and 
 \begin{equation}
  \begin{split}
    \tilde{j}_l(t) = \frac{1}{2l+1}\sum_{s=0}^{\infty}\, \beta_s^{(\hat{l}/2)}  \frac{(2l + 4s+1)!!}{(2s+l)!}\, \left(\frac{t}{1+t^2}\right)^{2s+l}\label{eq:ct8}
    \end{split}    
    \end{equation}
 are the corresponding modified spherical Bessel type functions. If one considers that ${t=\tan \alpha}$, and using the trigonometric relations ${1+\tan^2 \alpha = \sec^2 \alpha}$ and ${\sin 2\alpha =2\sin \alpha \cos \alpha}$,  the modified spherical Bessels type functions simplify to
 \begin{equation}
  \begin{split}
    \tilde{j}_l(\alpha) = \frac{1}{2l+1}\sum_{s=0}^{\infty}\, \beta_s^{(\hat{l}/2)}  \frac{(2l + 4s+1)!!}{(2s+l)!}\,  \left(\frac{1}{2}\sin 2 \alpha\right)^{2s+l}\label{eq:ct9}.
    \end{split}    
    \end{equation}
 The properties of the modified spherical Bessel type functions presented here need to be investigated further. Intuitively, we think that they belong to the family of the hyperspherical functions which usually have some recurrence relations.
 Equation (\ref{eq:ct9}) integrates the two electron co-ordinates as a correlated pair with ${r_1 =h \cos \alpha}$ and ${r_2 =h \sin \alpha}$ where $h$ is the distance between the two interacting electrons, equivalent to the hypotenuse of a right-angled triangle formed by the orthogonal vectors $\mathbf{r}_1$ and $\mathbf{r}_2$. 
 
 The first four orders of the modified spherical Bessel type functions, each with the first four terms of the expansion are:
 \begin{equation}
  \begin{split}
    \tilde{j}_0(\alpha) &=  1 + \frac{1}{3}\,\frac{5!!}{2!\,2^2}\, \sin^2(2\alpha) + \frac{1}{5}\,\frac{9!!}{4!\,2^4}\, \sin^4(2\alpha) \\& + \frac{1}{7}\,\frac{13!!}{6!\,2^6} \sin^6(2\alpha) + \cdots \\
    \tilde{j}_1(\alpha) &=  \frac{1}{2}\, \sin(2\alpha) + \frac{1}{5}\,\frac{7!!}{3!\,2^3}\, \sin^3(2\alpha) + \frac{1}{7}\,\frac{11!!}{5!\,2^5}\, \sin^5(2\alpha) \\ & + \frac{1}{9}\,\frac{15!!}{7!\,2^7}\, \sin^7(2\alpha) + \cdots \\
    \tilde{j}_2(\alpha) &=  \frac{2}{5\times 3}\, \frac{5!!}{2!\,2^2} \sin^2(2\alpha) + \frac{4}{7 \times 5}\, \frac{9!!}{4!\,2^4}\, \sin^4(2\alpha)\\ &+ \frac{6}{9 \times 7}\, \frac{13!!}{6!\,2^6}\, \sin^6(2\alpha) + \frac{8}{11 \times 9}\,\frac{17!!}{8!\,2^8}\, \sin^8(2\alpha) + \cdots \\
    \tilde{j}_3(\alpha) &= \frac{2}{7 \times 5}\, \frac{7!!}{3!\,2^3}\,\sin^3(2\alpha) + \frac{4}{9 \times 7}\,\frac{11!!}{5!\,2^5}\, \sin^5(2\alpha)\\ &+ \frac{6}{11 \times 9}\,\frac{15!!}{7!\,2^7}\, \sin^7(2\alpha) + \frac{8}{13 \times 11}\,\frac{19!!}{9!\,2^9}\, \sin^9(2\alpha) + \cdots  \label{eq:ct10}
    \end{split}    
    \end{equation}
  If one considers only the first term of each modified spherical Bessel  type functions, then the correlation term can be expressed as
 
 \begin{equation}
  \begin{split}
    \frac{1}{|\mathbf{r}_1-\mathbf{r}_2|}  &\approx \frac{\cos \alpha}{r_1}\, \sum_{l=0}  (2l+1)\,\left(\frac{1}{2}\sin 2 \alpha\right)^{l} P_l(\cos \theta)  \\
  &\approx  \frac{1}{\sqrt{r_1^2 + r_2^2}}\, \sum_{l=0} (2l+1)\, \left(\frac{r_1r_2}{r_1^2 + r_2^2}\right)^{l} P_l(\cos \theta)  \label{eq:ct11}.
    \end{split}    
    \end{equation}
 
 Our analytical expression in equation (\ref{eq:ct0}) is obtained from an intuitive consideration of this alternative multipole expansion series. The simplification using trigonometry in equation (\ref{eq:ct9}) implies that the two interacting electrons are mutually orthogonal to each other as expected from the principles of quantum mechanics. This geometry simplifies further the correlation interaction to 
 \begin{equation}
 \begin{split}
 \frac{1}{|\mathbf{r}_1-\mathbf{r}_2|} = \frac{1}{\sqrt{r_1^2 + r_2^2}} \label{eq:ct12}
 \end{split}
\end{equation}  
which needs to  be disentangled further. The proposed alternative multipole expansion, or any other method, can be used to approximate this coupled interaction while employing the fact that the vectors are orthogonal to each other in order to simplify the problem. Equation (\ref{eq:ct12}) is exactly similar to the hyperradius definition introduced by Macek \cite{Macek1968} in hyperspherical method. As opposed to the hyperspherical method in which the Hamiltonian of the two-electron system is expressed in terms of the hyperradius and the hyperangles \cite{Macek1968}, in this work we introduce separability of the Hamiltonian leading to an independent particle approximation solution to the Schr\"odinger equation but with the correlation effects fully embedded into the single electron Hamiltonian.
 
\section{Helium-like System  Pseudopotential}
Using the alternative multipole expansion, we developed the non-relativistic helium-like system pseudopotential
\begin{equation}
V(r) = -\frac{Z}{r} + \frac{1}{2} V_{\mathrm{scr}}(r,r')      \label{eq:pot1}
\end{equation}
for the independent particle Hamiltonian, where the first term is the interaction between the active electron and the nuclear charge $Z$, and $V_{\mathrm{scr}}$ is the central screening potential resulting from the other electron given by equation (\ref{eq:ct12}). Factor $1/2$ is based on the assumption that the correlation energy is shared equally between the two correlated electrons. This assumption should be accurate if the two electrons have identical quantum states (or identical principal quantum numbers). We have considered the two electrons to be indistinguishable, correlated, and likely to exchange their relative positions. 

By minimising the potential function in equation (\ref{eq:pot1}) by differentiating the function with respect to any of the radial co-ordinates and equating the derivative to zero yields the relation
 \begin{equation}
 \begin{split}
 \frac{1}{\sqrt{r_1^2 + r_2^2}} = \frac{\sqrt[3]{2Z}}{r_{1}} = \frac{\sqrt[3]{2Z}}{r_{2}} \label{eq:ct13}
 \end{split}
\end{equation}    
which introduces separability of the correlated term. We have used equation (\ref{eq:ct13}) as the screening potential in equation (\ref{eq:pot1}) to solve the time independent Schr\"odinger equation using an independent particle model
\begin{equation}
\begin{split}
\langle E \rangle =  \sum_{i=1,2} & [\langle \phi_{\beta}(\mathbf{r}_i)\mid \mathrm{H}_i \mid \phi_{\beta'}(\mathbf{r}_i) \rangle \\ &+ \langle \phi_{\beta}(\mathbf{r}_i)\mid \mathrm{H}_i \mid \phi_{\beta'}(\mathbf{r}_i) \rangle \delta_{\beta \beta'}] \label{eq:ip1}
\end{split}
\end{equation}
where the two-electron wavefunction has been expanded interms of the Slater-type orbitals and $\beta =\{n,l,m\}$ define the set of quantum numbers corresponding to any particular state. The first term of equation (\ref{eq:ip1}) emanates from the direct integral where no electron exchange is involved while the second term is the exchange integral which is non-vanishing only if $\beta = \beta'$. The interaction Hamiltonian $\mathrm{H}_n$ 
\begin{equation}
\mathrm{H}_n = \frac{1}{2}\,p_i^2 + V_n^{\mathrm{eff}}(\mathbf{r}_i,\mathbf{p}_i, \mathbf{s}_i) \label{eq:ham1}
\end{equation}
is defined for each independent electron with the index $n\geq 0$ taking integer values. The effective potential $V_n^{\mathrm{eff}}$ is a summation 
\begin{equation}
V_n^{\mathrm{eff}}(\mathbf{r}_i,\mathbf{p}_i, \mathbf{s}_i) = \sum_{i=0}^n V_{i}  \label{eq:pot2}
\end{equation} 
of some of the several terms of interaction drawn from equation ($39.14$) of Bethe and Salpeter \cite{amo:BetheandSalpeter1957}. Here we have explicitly mentioned and simplified further only the interactions that have been included in this work.  The first being the non-relativistic potential term $V_{0}$
\begin{equation}
V_{0}(\mathbf{r}_i) = -\frac{Z}{r_i} + \frac{\sqrt[3]{2Z + \chi_{\mathrm{corr}}(Z)}}{2\,r_i}, \label{eq:pot2a}
\end{equation}
 evaluated using equations (\ref{eq:pot1}) and (\ref{eq:ct13}) and it incorporates the electron correlation term. 
  The spin-spin interaction correction term $V_{1}$ can be simplified as 
\begin{equation}
\begin{split}
V_{1}(\mathbf{r}_i)& = \frac{\alpha^2}{2\,r_{ij}^3}\left(\mathbf{s}_i \cdot \mathbf{s}_j - \frac{3(\mathbf{s}_i \cdot \mathbf{r}_{ij})(\mathbf{s}_j \cdot \mathbf{r}_{ij})}{\mathbf{r}_{ij}^2} \right)\\
                    & = \frac{1}{2c^2}\,\frac{-2\,(\mathbf{s}_i \cdot \mathbf{s}_j)}{r_{ij}^3} \\
                    & \approx \frac{[2Z + \chi_{\mathrm{corr}}(Z)]}{(2c)^2\,r_i^3} \label{eq:pot2b}
\end{split}
\end{equation}
having used equation (\ref{eq:ct13}) and where $c$ is the reciprocal of the fine structure constant ($\alpha$). Considering the current definition of the electron correlation term, the first term of this spin-spin interaction as defined in equation (39.14) of Bethe and Salpeter \cite{amo:BetheandSalpeter1957} vanishes because of the boundary conditions of the wavefunction and the Dirac delta condition. The approximation in equation (\ref{eq:pot2b}) is based on making a classical argument that $(\mathbf{s}_i \cdot \mathbf{s}_j)$ is equal to $-1/4$ instead of the quantum mechanical prescribed value of  $-3/4$ for the singlet states. This is equivalent to considering only a third of singlet spin-spin interaction term value because the spins are assumed to be aligned parallel or antiparallel to one particular direction.

The term $V_4$
    \begin{equation}
    \begin{split}
    V_{2}(\mathbf{r}_i)& =\frac{1}{2} \frac{1}{(2c)^2}\, \nabla \cdot (-\nabla V(\mathbf{r}_i)) \\
                       & = -\frac{1}{2} \frac{1}{(2c)^2}\,\frac{\partial ^2 V(r_i)}{\partial r_i^2} \\
                       & = \frac{1}{(2c)^2}\,\left( \frac{Z}{r_i^3} - \frac{\sqrt[3]{2Z + \chi_{\mathrm{corr}}(Z)}}{2\,r_i^3} \right)
    \end{split}
    \end{equation}
    is a characteristic of the Dirac theory with the potential function $V(\mathbf{r}_i)$ already defined in equation (\ref{eq:pot1}).
 The classical relativistic correction $V_{3}$ 
 \begin{equation}
 \begin{split}
 V_{3}(\mathbf{r}_{12}, \mathbf{p}_{i,j}) &= -\frac{1}{2\,c^2}\, \frac{1}{r_{12}} \left(\mathbf{p}_i \cdot \mathbf{p}_j + \frac{\mathbf{r}_{ij}\cdot (\mathbf{r}_{ij} \cdot \mathbf{p}_{i})\mathbf{p}_j }{\mathbf{r}_{ij}} \right)\\
        &= -\frac{1}{2c^2}\, \frac{1}{r_{12}}\,[2\,(\mathbf{p}_i \cdot \mathbf{p}_j)]\\
        &= -\frac{1}{2c^2}\, \frac{1}{r_{12}}\,[p_i^2 + p_j^2 - P^2]\\
        &= -\frac{1}{2c^2}\, \frac{\sqrt[3]{2Z + \chi_{\mathrm{corr}}(Z)}}{r_{i}}\,[p_i^2 + p_j^2 - P^2]
 \end{split}
 \end{equation}
 to the interaction between electrons. Here $\mathbf{P} = \mid \mathbf{p}_i - \mathbf{p}_j \mid $ vanishes if $i=j$. This term reduces to
    \begin{equation}
    V_{3}(\mathbf{r}_{i}, \mathbf{p}_{i})  = -\frac{1}{2c^2}\, \frac{\sqrt[3]{2Z + \chi_{\mathrm{corr}}(Z)}}{r_{i}}\,p_i^2
    \end{equation}
  if it is separated for each of the individual electron co-ordinates.
  The finite mass correction term $V_{4}$
 \begin{equation}
V_{4}(\mathbf{r}_i) = -\frac{1}{M}\, \mathrm{H}_{\infty} \label{eq:pot2c}
\end{equation}
 has been obtained from reference \cite{amo:Bransden1990} with $\mathrm{H}_{\infty}$ as the Hamiltonian of the system without the finite mass correction,  ${1/M}$ is the electron-nucleon mass ratio. 
  The scalar function $\chi_{\mathrm{corr}}$ in $V_{n=0,1,2,3}$ 
\begin{equation}
\chi_{\mathrm{corr}}(Z) = \gamma^Z\,Z\,(Z-2) \label{eq:pot3}
\end{equation}
is a fitting function optimized to offer the additional correction $V_{5}$ for the ionic systems but vanishes for the helium atom. The adjustable parameter $\gamma=1.0821$ yields good quantitative agreement with experimental results for the groundstate energies of ionic systems investigated.

We have used the Hamiltonian as defined in equation (\ref{eq:ham1}) and diagonalized it in a B spline spectral basis set having a box radius of $400$ au, $1200$ B splines of order $k=10$, and a non-linear knot sequence. As already stated, the goal was to test efficiency of the present method proposed in this work. With the analytical expression of the electron correlation term, it was also found desirable to include some corrections to the Schr\"odinger equation for two-electron systems that could be evaluated without further complexities. The inclusion of the correction terms also show the relative importance of the additional interactions as compared to the non-relativistic terms.

The non-relativistic eigenvalue for the groundstate  energy of helium resulting from this method is in good agreement with the experimental value as shown in table \ref{tab1}. Furthermore, the discrepancy between the experimental ground state potential and the obtained theoretical non-relativistic value is properly accounted for by including some of the correction terms like spin-spin coupling, classical relativistic correction, the characteristic Dirac theory term, and the finite mass correction term. We can therefore consider the theoretical value $-2.9103$ from our calculations to be the correct non-relativistic threshold groundstate energy for helium atom. The very accurate  groundstate energy as calculated by the Hylleraas method \cite{Drake1999}, from this hypothesis, includes all the corrections beyond the non-relativistic energy. This explanation may be justified based on the fact that the accurate value obtained using the Hylleraas method is very close to the experimentally obtained values of Bergeson \textit{et. al.}\cite{Bergeson1998} and Eikama \textit{et. al.} \cite{Eikama1997}. The experimental values are expected to incorporate all orders of correction beyond the non-relativistic Hamiltonian to the groundstate energy value, including all QED and finite mass corrections. The method adopted in this work, if ascertained to be valid, can be a great numerical feat emanating from the use of perturbative methods to account for the most significant terms responsible for the groundstate energy of helium atom without using any adjustable parameters. 

We have also determined the excitation energies of the $2s^2$ and $2p^2$ autoionizing states from this method to be $59.22$ eV and $59.20$ eV respectively against the known experimental values of $57.8$ eV and $62.2$ eV \cite{Rudd1964} respectively. Although the present method seems to be almost exact for the groundstate eigenvalue for helium, the discrepancy between the theoretical and the experimental values of the $2s^2$ and $2p^2$ singlet autoionizing states shows that corrections included may still not be sufficient for accurate description of these states.

\begin{table}[!ht]
    \centering
    \begin{tabular}{lllllll}
    \hline
    State & $H_{0}$ & $H_{1}$  &  $H_{2}$ & $H_{3}$ & $H_{4}$ & Exp.t  \\
   \hline
   \hline
       $1s^2$& -2.9103  & -2.8996 &-2.8968 & -2.9040 & -2.9036 &  -2.9037    \\                                   
%
     $2s^2$& -0.7276     & -0.7263 & -0.7259  & -0.7268 & -0.7267 & -0.7787    \\                          
%
    $2p^2$&   -0.7276   & -0.7276 &-0.7276  & -0.7276  &-0.7275 & -0.6169   \\                                   
    \hline
    \end{tabular}
    \caption{Some numerically calculated eigenvalues using the present model potential versus the reference experimental values for helium groundstate \cite{Bergeson1998, Eikama1997} and autoionizing levels \cite{Rudd1964}.  The ${H_{0}=\frac{1}{2}\,p_i^2 + V_0}$ represents the theoretical non-relativistic Hamiltonian, ${H_{n=1,2, \cdots}= H_0 + \sum_{i=1}^n V_i}$ is the effective Hamiltonian including correction terms $V_i$  already defined in equations (\ref{eq:pot2a})-(\ref{eq:pot2c}).}
    \label{tab1}
  \end{table}

We extended the method to other two-electron systems for $1\le Z \le 6$ nuclear charges. Table \ref{tab2} shows the groundstate energies for the two-electron systems corresponding to the present non-relativistic model and the extra corrections outlined. The additional $H_5$ data is obtained if an additional confinement is introduced by the fitting function defined in equation (\ref{eq:pot3}).  
\begin{table}[!ht]
    \centering
    \begin{tabular}{llllllll}
    \hline
   $Z$ & $H_{0}$ & $H_{1}$  &  $H_{2}$ & $H_{3}$ & $H_{4}$& $H_{5}$ & Exact  \\
   \hline
   \hline
       $1$& -0.2739  & -0.2737 & -0.2736 & -0.2738 &  -0.2737 & -0.5285 &-0.528     \\                                   
%
     $2$& -2.9103     & -2.8996 & -2.8968  & -2.9040 &  -2.9036 & -2.9036 &-2.9037     \\                          
%
    $3$&   -8.7482   & -8.6756 &-8.6555  & -8.6966  & -8.6959 & -7.3794 & -7.28    \\
     
     $4$& -18.000  & -17.746 &-17.675 & -17.803  & -17.802 & -13.913 &-13.66    \\                                   
%
     $5$& -30.776     & -30.137 & -29.963  & -30.257 & -30.255 & -22.340 & -22.03     \\                          
%
    $6$&   -47.148   & -45.825 &-45.474  &-46.041   & -46.040 & -32.413 & -32.41   \\                                  
    \hline
    \end{tabular}
    \caption{Similar to table \ref{tab1} but for groundstate eigenvalues of helium-like systems. All the columns, except the additional $H_{5}$ column, take a zero value for the fitting function defined in equation (\ref{eq:pot3}). The exact values have been extracted from ref. \cite{amo:Bransden1990}.}
    \label{tab2}
  \end{table}
  
From table \ref{tab2}, one can observe that there is a systematic deviation of the present results from the exact experimental values of the groundstate energies of the ionic systems despite its success with the helium atom. However, if the present model is applied with the additional correction introduced by the fitting function defined in equation (\ref{eq:pot3}), quite a good agreement with the expected results is achieved. This seems to suggest that there is an additional potential present in the ionic species due to the net charge in the system, but absent in the neutral atom.
  
\section{Conclusion}
We have developed an alternative multipole expansion of the electron-electron correlation term which suggests that the two interacting electrons are mutually perpendicular to each other. This simplifies the interaction term making the Schr\"odinger equation separable for each of the two-electron co-ordinates. We use this separability to obtain a non-relativistic threshold energy of the helium atom in its groundstate. We also show perturbatively that the experimental ground state energy value includes additional higher order corrections to the calculated non-relativistic energy. 

The classical relativistic corrections and the spin-spin coupling offer the most dominant corrections to the non-relativistic limit. Furthermore, the present method predicts a systematic deviation of the calculated non-relativistic groundstate energies of the two-electron ions relative to the experimental values despite its success with the helium atom. A slight modification to the derived electron correlation term is intuitively introduced to account for this discrepancy. If the present method is justified, the discrepancy in the ionic helium-like systems suggest that there is an additional interactions, due to the charge surplus in the system, not accounted for by the known corrections to the two-electron problem.

Despite the success of the proposed method with the groundstate energy of helium atom, the large deviations for the helium-like ions as well as the autoionizing levels warrant further investigation. One can also see the possibility of improving this method further as a solution to the many-body problem. 

\section{Acknowledgement}
We are grateful to NACOSTI and DAAD for funding this project, and to AG Moderne Optik of Humboldt Universit\"at zu Berlin for providing the computational resources used in this work.

\bibliographystyle{apsrev}
\bibliography{/home/eric/Inworks/Literature}

\end{document}